\documentclass[aps,reprint,amsmath,amssymb,prb,superscriptaddress,floatfix,pdftex]{revtex4-1}
\usepackage[pdftex]{graphicx}
\usepackage{dcolumn}
\usepackage{bm}
\usepackage{braket}
\usepackage{color}

\allowdisplaybreaks[4] 

\begin{document}


\title{Benchmark of density functional theory for superconductors in elemental materials}



\author{Mitsuaki Kawamura}
\email{mkawamura@issp.u-tokyo.ac.jp}
\affiliation{Institute for Solid State Physics, 
  The University of Tokyo, Kashiwa 277-8581, Japan}

\author{Yuma Hizume}
\affiliation{Department of Physics, 
  The University of Tokyo, Tokyo 113-0033, Japan}

\author{Taisuke Ozaki}
\affiliation{Institute for Solid State Physics, 
  The University of Tokyo, Kashiwa 277-8581, Japan}

\email[]{mkawamura@issp.u-tokyo.ac.jp}
\homepage[]{https://mitsuaki1987.github.io/}


\date{\today}

\begin{abstract}
  Systematic benchmark calculations for elemental bulks are presented to validate the accuracy of density functional theory for superconductors.
  We developed a method to treat the spin-orbit interaction (SOI) together with the spin fluctuation (SF) and examine their effect on the superconducting transition temperature.
  We found the following results from the benchmark calculations:
  (1) The calculations, including SOI and SF, reproduce the experimental superconducting transition temperature ($T_c$) quantitatively.
  (2) The effect by SOI is small excepting a few elements such as Pb, Tl, and Re.
  (3) SF reduces $T_c$s, especially for the transition metals,
  while this reduction is too weak to reproduce the $T_c$s of Zn and Cd.
  (4) We reproduced the absence of superconductivity for alkaline (earth) and noble metals.
  These calculations confirm that our method can be applied to a wide range of materials and implies a direction for the further improvement of the methodology.
\end{abstract}

\pacs{}

\maketitle

\section{Introduction}

The first-principles calculation of the superconducting properties
such as the transition temperature ($T_c$) and the gap function is of great interest
to explore new materials as well as to understand the physical mechanism of known superconductors.
Density functional theory for superconductors (SCDFT)
\cite{PhysRevLett.60.2430,PhysRevB.72.024545}
is one of the frameworks
for such calculations;
this method enables us to perform fully non-empirical simulations
in the superconducting phases at a reasonable computational cost.
The anisotropic Migdal-Eliashberg (ME) equations \cite{PhysRevB.87.024505}
and the McMillan's formula \cite{PhysRev.167.331,Dynes1972615}
which is the parametrization of the solution of the ME equations
can also be used to estimate $T_c$.
However, to solve the Migdal-Eliashberg equations\cite{doi:10.7566/JPSJ.87.041012}, we need to perform
the summation of the Matsubara frequencies and this summation requires a substantial computational cost.
Since the McMillan's formula involves an adjustable parameter to
evaluate the effect of the Coulomb repulsion,
the formula cannot compare the $T_c$s of a wide range of materials.
In SCDFT, we can treat the electron-phonon interaction,
the electron-electron repulsion,
and the spin-fluctuation (SF)-mediated interaction \cite{PhysRevB.90.214504}
in a first-principles manner.
SCDFT has been applied to various kinds of materials such as
elemental materials (Al, Nb, Mo, Ta, Pb) \cite{PhysRevB.72.024546},
MgB$_2$ \cite{PhysRevLett.94.037004},
graphite intercalations \cite{PhysRevB.75.020511},
Li under high pressure \cite{PhysRevLett.96.047003},
H$_2$ molecule solid \cite{PhysRevLett.100.257001},
hydrogen compounds \cite{PhysRevB.91.224513}, and
FeSe \cite{PhysRevB.94.014503}.
On the other hand, the methodological improvements have also been proposed to include
the anisotropic electron-phonon interaction,
plasmons \cite{PhysRevLett.111.057006},
spin-fluctuation \cite{PhysRevB.90.214504},
and the spin-orbit interaction (SOI) \cite{PhysRevB.101.014505}.

However, the accuracy of the current approximated functional of SCDFT
and the effects of SOI and SF
have not been verified systematically
although such verification is highly desirable
before applying this method to a wide range of materials.
Such a high-throughput calculation was performed,
for example, in the exploration of low-thermal-conductivity compounds using first-principles calculations together with the
materials informatics \cite{PhysRevLett.115.205901}.
A benchmark is also a useful tool used to find a guideline
for improving the theory and approximations of the superconducting density functional.
For this purpose, in this paper, we are presenting the benchmark calculations of SCDFT.
As benchmark targets, we have chosen the simplest superconducting and
non-superconducting materials, i.e., elemental materials;
each material in this group comprises a single element.
The particular computational cost is relatively low because most materials in this group contain only one or two atoms in the unit cell.
Moreover, we can see the effects of the chemical difference and the strength of the SOI of each element.

This paper is organized as follows:
In Sec.~\ref{sec_theory} we explain the theoretical foundations of SCDFT, including SF and SOI,
and in Sec.~\ref{sec_implement}, details of the mathematical formulation and the implementation are shown.
Next, we list the results together with the numerical condition in Sec.~\ref{sec_result}, and present the study discussion in Sec.~\ref{sec_discussion}.
Finally, we summarize the study in Sec.~\ref{sec_summary}.

\section{Theory} \label{sec_theory}

In this section, we will explain in detail the SCDFT formulation
including plasmon-aided mechanism \cite{PhysRevLett.111.057006},
SF effect \cite{PhysRevB.90.214504}, and
SOI \cite{PhysRevB.101.014505}.
We use the Hartree atomic units throughout the paper.
In this study, we only consider the singlet superconductivity,
while in Ref. \onlinecite{PhysRevB.101.014505}, both the singlet- and triplet-superconducting states were considered.
Within SCDFT, $T_c$ is obtained as a temperature where the following
Kohn-Sham superconducting gap $\Delta_{n {\bf k}}$ becomes zero at all
the band $n$ and wavenumber ${\bf k}$:
\begin{align}
  \Delta_{n {\bf k}} &= -\frac{1}{2} \sum_{n' {\bf k}'}
  \frac{K_{n {\bf k} n'{\bf k}'}(\xi_{n{\bf k}},\xi_{n'{\bf k}'})}
       {1+Z_{n {\bf k}}(\xi_{n{\bf k}})}
       \nonumber \\
       &\times \frac{\Delta_{n' {\bf k}'}}
              {\sqrt{\xi_{n' {\bf k}'}^2 + \Delta^2_{n' {\bf k}'}}}
       \tanh\left(\frac{\sqrt{\xi_{n' {\bf k}'}^2 + \Delta^2_{n' {\bf k}'}}}{2 T}\right),
  \label{eq_gapeq}
\end{align}
where $\xi_{n {\bf k}}$ is the Kohn-Sham eigenvalue measured from the Fermi level ($\varepsilon_{\rm F}$)
at the band index $n$ and wave-number ${\bf k}$.
$\xi_{n {\bf k}}$ is obtained by solving the following spinor Kohn-Sham equation:
\begin{align}
  &
  \begin{pmatrix}
    -\frac{\nabla^2}{2} + V^{\rm KS}_{\uparrow \uparrow}({\bf r}) - \varepsilon_{\rm F} &
    V^{\rm KS}_{\uparrow \downarrow}({\bf r}) \\
    V^{\rm KS}_{\downarrow \uparrow}({\bf r}) &
    -\frac{\nabla^2}{2} + V^{\rm KS}_{\downarrow \downarrow}({\bf r}) - \varepsilon_{\rm F}
  \end{pmatrix}
  \begin{pmatrix}
    \varphi_{n {\bf k} \uparrow}({\bf r}) \\ \varphi_{n {\bf k} \downarrow}({\bf r})
  \end{pmatrix}
  \nonumber \\
  &\hspace{1cm}=
  \xi_{n {\bf k}}
  \begin{pmatrix}
    \varphi_{n {\bf k} \uparrow}({\bf r}) \\ \varphi_{n {\bf k} \downarrow}({\bf r})
  \end{pmatrix},
\end{align}
where $\varphi_{n {\bf k} \sigma}({\bf r})$ is the $\sigma$ component of the spinor Kohn-Sham orbital
at $(n, {\bf k})$, and
$V^{\rm KS}_{\sigma \sigma'}({\bf r})$ is the $\sigma \sigma'$ component of the Kohn-Sham potential
with SOI ($\sigma, \sigma' = \uparrow, \downarrow$).
Due to the off-diagonal part of the Kohn-Sham potential,
the spin-up state and the spin-down state are hybridized.
Therefore, the Kohn-Sham eigenvalue $\xi_{n {\bf k}}$ does not have a spin index ($\sigma$).
The non-linear gap equation (\ref{eq_gapeq}) should be solved numerically
at each temperature.
The integration kernel $K_{n {\bf k} n' {\bf k}'}(\xi,\xi')$ indicates
the superconducting-pair breaking and creating interaction and
comprises the following three terms:
\begin{align}
  K_{n {\bf k} n' {\bf k}'} (\xi,\xi') &\equiv
  K^{ep}_{n {\bf k} n'{\bf k}'}(\xi,\xi')
  + K^{ee}_{n {\bf k} n' {\bf k}'}(\xi,\xi')
  \nonumber \\
  &+ K^{sf}_{n {\bf k} n'{\bf k}'}(\xi,\xi'),
  \label{eq_k_all}
\end{align}
namely, the electron-phonon, Coulomb repulsion,
and spin-fluctuation kernel, respectively.
However,
the renormalization factor $Z_{n {\bf k}}(\xi_{n{\bf k}})$ comprises
only the electron-phonon and spin-fluctuation terms as follows.
\begin{align}
  Z_{n {\bf k}}(\xi) \equiv
  Z_{n {\bf k}}^{ep}(\xi)+Z_{n {\bf k}}^{sf}(\xi),
  \label{eq_z_all}
\end{align}
because the Coulomb-repulsion contribution to this factor is
already included in the Kohn-Sham eigenvalue $\xi_{n {\bf k}}$.
The temperature $T$ is defined by considering the Boltzmann constant $k_{\rm B}=1$.

Let us explain each term in the kernel and the renormalization factor below.
The electron-phonon kernel $K^{ep}$ and renormalization factor $Z^{ep}$ are given by
\cite{PhysRevB.101.014505}
\begin{align}
  &K_{n {\bf k} n' {\bf k}'}^{ep}(\xi,\xi') =
  \frac{2}{\tanh[\xi/(2T)] \tanh[\xi'/(2T)]}
  \sum_{\nu}
  |g_{n {\bf k} n' {\bf k}'}^{\nu}|^2
  \nonumber \\
  &\hspace{0.3cm}
  \times[I(\xi, \xi', \omega_{{\bf k}'-{\bf k} \nu})
  - I(\xi, -\xi', \omega_{{\bf k}'-{\bf k} \nu})],
  \label{eq_kep}
  \\
  &Z_{n {\bf k}}^{ep}(\xi) =
  \frac{-1}{\tanh[\xi/(2T)]} \sum_{n' {\bf k}' \nu}
  |g_{n {\bf k} n' {\bf k}'}^{\nu}|^2
  \nonumber \\
  &\hspace{0.5cm}
  \times[J(\xi, \xi_{n' {\bf k}'}, \omega_{{\bf k}'-{\bf k} \nu})
  + J(\xi, -\xi_{n' {\bf k}'}, \omega_{{\bf k}'-{\bf k} \nu})],
  \label{eq_zep}
\end{align}
where $\omega_{{\bf q} \nu}$ is the phonon frequency at wave-number ${\bf q}$
and branch $\nu$.
$I(\xi,\xi',\omega)$ and $J(\xi,\xi',\omega)$ are derived with the Kohn-Sham perturbation theory \cite{PhysRevA.50.196}, and are written as follows \cite{PhysRevB.72.024545}:
\begin{align}
  I(\xi,\xi',\omega) &= f_T(\xi)f_T(\xi')n_T(\omega)
  \nonumber \\
  &\hspace{-1.3cm}\times \left[
    \frac{e^{\xi/T}-e^{(\xi'+\omega)/T}}{\xi-\xi'-\omega}
    -  \frac{e^{\xi'/T}-e^{(\xi+\omega)/T}}{\xi-\xi'+\omega}
    \right],
  \\
  J(\xi,\xi',\omega) &= {\tilde J}(\xi,\xi',\omega) - {\tilde J}(\xi,\xi',\omega),
  \\
    {\tilde J}(\xi,\xi',\omega) &=
    -\frac{f_T(\xi)+n_T(\xi)}{\xi-\xi'-\omega}
    \nonumber \\
    & \hspace{-1.3cm}\times \left[
      \frac{f_T(\xi')-f_T(\xi-\omega)}{\xi-\xi'-\omega}
      - \frac{f_T(\xi-\omega) f_T(-\xi'+\omega)}{T}
      \right],
\end{align}
where $f_T(\xi)$ and $n_T(\omega)$ are the Fermi-Dirac and the Bose-Einstein
distribution function, respectively.
The functions $I(\xi,\xi',\omega)$ and $J(\xi,\xi',\omega)$ yield a temperature-dependent retardation effect.
The electron-phonon vertex $g$ between
Kohn-Sham orbitals indexed with $(n,{\bf k})$ and $(n',{\bf k}+{\bf q})$, and
the phonon $({\bf q},\nu)$ is computed as \cite{PhysRevB.81.174527}
\begin{align}
  g_{n {\bf k} n' {\bf k}+{\bf q}}^{\nu} &= \int d^3 r 
  \sum_{\sigma \sigma'=\uparrow,\downarrow} \varphi_{n' {\bf k}+{\bf q} \sigma}^*({\bf r}) \varphi_{n {\bf k} \sigma'}({\bf r})
  \nonumber \\
  &\times
  \sum_{\tau}
  \frac{{\boldsymbol \eta}_{{\bf q} \nu}^{\tau} \cdot
    {\boldsymbol \delta}^{{\bf q}\tau} V^{\rm KS}_{\sigma \sigma'}({\bf r})
  }{\sqrt{2 M_\tau \omega_{{\bf q} \nu}}},
  \label{eq_gep}
\end{align}
where $M_\tau$ is the mass of atom labeled by $\tau$,
${\boldsymbol \eta}_{{\bf q} \nu}^\tau$ is the polarization vector of phonon $({\bf q}, \nu)$ and atom $\tau$,
and
${\boldsymbol \delta}^{{\bf q}\tau} V^{\rm KS}_{\sigma \sigma'}({\bf r})$
is the Kohn-Sham potential deformed by the periodic displacement of atom $\tau$ and wave number ${\bf q}$
\begin{align}
  {\boldsymbol \delta}^{{\bf q}\tau} V^{\rm KS}_{\sigma \sigma'}({\bf r})
  =\sum_{\bf R} e^{i {\bf q}\cdot{\bf R}}
  \frac{\delta V^{\rm KS}_{\sigma \sigma'}[\{{\bf r}_{\tau {\bf R}}\}]({\bf r})}
  {\delta {\bf r}_{\tau {\bf R}}},
\end{align}
where ${\bf r}_{\tau {\bf R} }$ is the position of the atom $\tau$ at the cell ${\bf R}$. 
We have obtained the deformation potential by the phonon calculation,
based on density functional perturbation theory (DFPT) \cite{RevModPhys.73.515}.
The electron-phonon kernel $K^{ep}$ is always negative; therefore, it makes a positive contribution in forming the Cooper pair.
However, the electron-phonon renormalization factor $Z^{ep}$ weakens the effect caused by the kernels.

The electron-electron repulsion kernel $K^{ee}$ in Eq.~(\ref{eq_k_all}) is \cite{PhysRevB.101.014505}
\begin{align}
  K_{n {\bf k} n' {\bf k}'}^{ee}(\xi,\xi') =
  \frac{2}{\pi}\int_0^\infty d\omega
  \frac{|\xi|+|\xi'|}
       {(|\xi|+|\xi'|)^2+\omega^2}
  V_{n {\bf k} n' {\bf k}'}^{ee} (i\omega),
  \label{eq_kee}
\end{align}
where $V_{n {\bf k} n' {\bf k}'}^{ee} (i\omega)$ is the dynamically screened exchange integral between
the Kohn-Sham orbitals $(n, {\bf k})$ and $(n', {\bf k}')$
\begin{align}
  V_{n {\bf k} n' {\bf k}'}^{ee} (i\omega)
  &= \iint d^3 r d^3 r' 
  V_{RPA}({\bf r}, {\bf r}', i\omega)
  \nonumber \\
  &\hspace{1cm} \times
  \rho^{(0)}_{n {\bf k} n' {\bf k}'}({\bf r}) \rho^{(0)*}_{n {\bf k} n' {\bf k}'}({\bf r}'),
  \label{eq_vee}
  \\
  \rho^{(0)}_{n {\bf k} n' {\bf k}'}({\bf r}) &=
  \sum_{\sigma = \uparrow, \downarrow} \varphi_{n {\bf k} \sigma}^*({\bf r}) \varphi_{n' {\bf k}' \sigma}({\bf r}).
  \label{eq_rho0}
\end{align}
In this study, we have computed the screened Coulomb interaction $V_{RPA}$ by applying the random phase approximation (RPA) \cite{PhysRev.106.364} as
\begin{align}
  &V_{RPA}({\bf r}, {\bf r}', i\omega) = \frac{1}{|{\bf r} - {\bf r}'|}
  + \iint d^3r_1 d^3r_2
  \nonumber \\
  &\hspace{1cm} \times
  V_{RPA}({\bf r}, {\bf r}_1, i\omega) \Pi_{\rm KS}^{00}({\bf r}_1, {\bf r}_2, i\omega)
  \frac{1}{|{\bf r}_2 - {\bf r}'|},
  \label{eq_vrpa}
\end{align}
where $\Pi_{\rm KS}^{0 0}$ is the electronic susceptibility of the Kohn-Sham system (the non-perturbed susceptibility).
This electronic susceptibility is the $\alpha=0$ part of the following susceptibilities of the Kohn-Sham system:
\begin{align}
  \Pi_{\rm KS}^{\alpha \alpha}({\bf r}, {\bf r}',i\omega) = \sum_{{\bf k} {\bf k}' n n'} &
  \frac{\theta(-\xi_{n {\bf k}})
    - \theta(-\xi_{n' {\bf k}'})}
       {\xi_{n {\bf k}}-\xi_{n' {\bf k}'}+i\omega}
  \nonumber \\
  &\times
  \rho^{(\alpha)}_{n {\bf k} n' {\bf k}'}({\bf r}) \rho^{(\alpha)*}_{n {\bf k} n' {\bf k}'}({\bf r}'),
  \label{eq_piks}
\end{align}
where $\alpha$ takes 0, $x$, $y$, $z$, and
\begin{align}
  \rho^{(x)}_{n {\bf k} n' {\bf k}'}({\bf r}) &=
  \sum_{\sigma=\uparrow, \downarrow}
  \varphi_{n {\bf k} \sigma}^*({\bf r}) \varphi_{n' {\bf k}' -\sigma}({\bf r}),
  \label{eq_rhox}
  \\
  \rho^{(y)}_{n {\bf k} n' {\bf k}'}({\bf r}) &=
  \sum_{\sigma=\uparrow, \downarrow}
  \sigma \varphi_{n {\bf k} \sigma}^*({\bf r}) \varphi_{n' {\bf k}' -\sigma}({\bf r}),
  \label{eq_rhoy}
  \\
  \rho^{(z)}_{n {\bf k} n' {\bf k}'}({\bf r}) &=
  \sum_{\sigma=\uparrow, \downarrow}
  \sigma \varphi_{n {\bf k} \sigma}^*({\bf r}) \varphi_{n' {\bf k}' \sigma}({\bf r}).
  \label{eq_rhoz}
\end{align}
The spin susceptibility $\Pi_{\rm KS}^{xx}$, $\Pi_{\rm KS}^{yy}$, and
$\Pi_{\rm KS}^{zz}$ are
used in the spin-fluctuation term later on.
Because of the factor
$[\theta(-\xi_{n {\bf k}}) - \theta(-\xi_{n' {\bf k}'})]/(\xi_{n {\bf k}}-\xi_{n' {\bf k}'}+i\omega)$,
these susceptibilities are affected largely by the electronic states in the vicinity of Fermi surfaces.

We propose the spin-fluctuation (SF) kernel $K^{sf}$ in Eq.~(\ref{eq_k_all}) and the renormalization $Z^{sf}$ in Eq.~(\ref{eq_z_all}) constructed using the noncollinear spinor wavefunctions.
The following formulation is an extension of those quantities in the collinear magnetism \cite{PhysRevB.94.014503}.
\begin{align}
  K_{n {\bf k} n' {\bf k}'}^{sf}(\xi,\xi') &=
  \frac{2}{\pi}\int_0^\infty d\omega
  \frac{|\xi|+|\xi'|}{(|\xi|+|\xi'|)^2+\omega^2}
  \nonumber \\
  & \times\Lambda_{n {\bf k} n' {\bf k}'}^{sf}(i\omega),
  \label{eq_ksf}
  \\
  Z_{n {\bf k}}^{sf}(\xi) &=
  \frac{1}{\pi} \sum_{n'{\bf k}'} \int_0^\infty d\omega
  \frac{(|\xi|+|\xi_{n' {\bf k}'}|)^2 - \omega^2}
       {[ (|\xi|+|\xi_{n' {\bf k}'}|)^2+\omega^2 ]^2}
  \nonumber \\
  &\times \Lambda_{n {\bf k} n' {\bf k}'}^{sf}(i\omega),
  \label{eq_zsf}
\end{align}
where
\begin{align}
  \Lambda_{n {\bf k} n' {\bf k}'}^{sf}(i\omega)
  &= \sum_{\alpha=x,y,z} \iint d^3 r d^3 r' 
  \Lambda^{sf}_{\alpha \alpha}({\bf r}, {\bf r}', i\omega)
  \nonumber \\
  &\hspace{1cm} \times
  \rho^{(\alpha)}_{n {\bf k} n' {\bf k}'}({\bf r})
  \rho^{(\alpha)*}_{n {\bf k} n' {\bf k}'}({\bf r}').
  \label{eq_lambdasf}
\end{align}
$\Lambda_{n {\bf k} n' {\bf k}'}^{sf}$ has a similar form to
the screened exchange integral of Eq.~(\ref{eq_vee}),
and it involves the summation over the $x, y$, and $z$ components
of the following SF-mediated interaction:
\begin{align}
  &\Lambda^{sf}_{\alpha \alpha}({\bf r}, {\bf r}',i\omega) = - \iint d^3 r_1 d^3 r_2
  \nonumber \\ 
  &\hspace{1cm}\times I_{XC}^{\alpha \alpha}({\bf r}, {\bf r}_1)
  \Pi^{\alpha \alpha}({\bf r}_1, {\bf r}_2, i\omega)
  I_{XC}^{\alpha \alpha}({\bf r}_2, {\bf r}'),
  \label{eq_hatlambdasf}
\end{align}
where $\Pi^{\alpha \alpha}$ is the spin susceptibility of the interacting system as \cite{PhysRevLett.55.2850}
\begin{align}
  &\Pi^{\alpha \alpha}({\bf r}, {\bf r}', i\omega)
  = \Pi_{\rm KS}^{\alpha \alpha}({\bf r}, {\bf r}', i\omega)
  +\iint d^3 r_1 d^3 r_2
  \nonumber \\
  &\hspace{1cm} \times
  \Pi^{\alpha \alpha}({\bf r}, {\bf r}_1, i\omega)
  I_{XC}^{\alpha \alpha}({\bf r}_1, {\bf r}_2)
  \Pi_{\rm KS}^{\alpha \alpha}({\bf r}_2, {\bf r}', i\omega).
  \label{eq_pi_int}
\end{align}
In Eqs. (\ref{eq_hatlambdasf}) and (\ref{eq_pi_int}),
the spin-spin interaction is included through the exchange correlation kernel:
\begin{align}
  I_{XC}^{\alpha \alpha} ({\bf r}, {\bf r}')
  \equiv \frac{\delta^2 E_{XC}}{\delta m_\alpha({\bf r}) \delta m_\alpha({\bf r}')}
  \label{eq_xckernel}
\end{align}
which is the second-order functional derivative of the exchange-correlation energy $E_{XC}$
with respect to the spin density along the $\alpha$ direction, $m_\alpha$.
We have used the results of the standard density functional calculations of the normal (non-superconducting) state to calculate the abovementioned quantities.
Therefore, we have computed $T_c$ by solving the gap equation (\ref{eq_gapeq})
as a post-process of the calculations of the normal state.
The treatment described, known as the decoupling approximation, is known to be reliable
when the bandwidth and the superconducting gap energy scales are largely different
\cite{PhysRevB.72.024545}.

\section{Implementation} \label{sec_implement}

\begin{figure*}[!tb]
  \begin{center}
    \includegraphics[width=17.9cm]{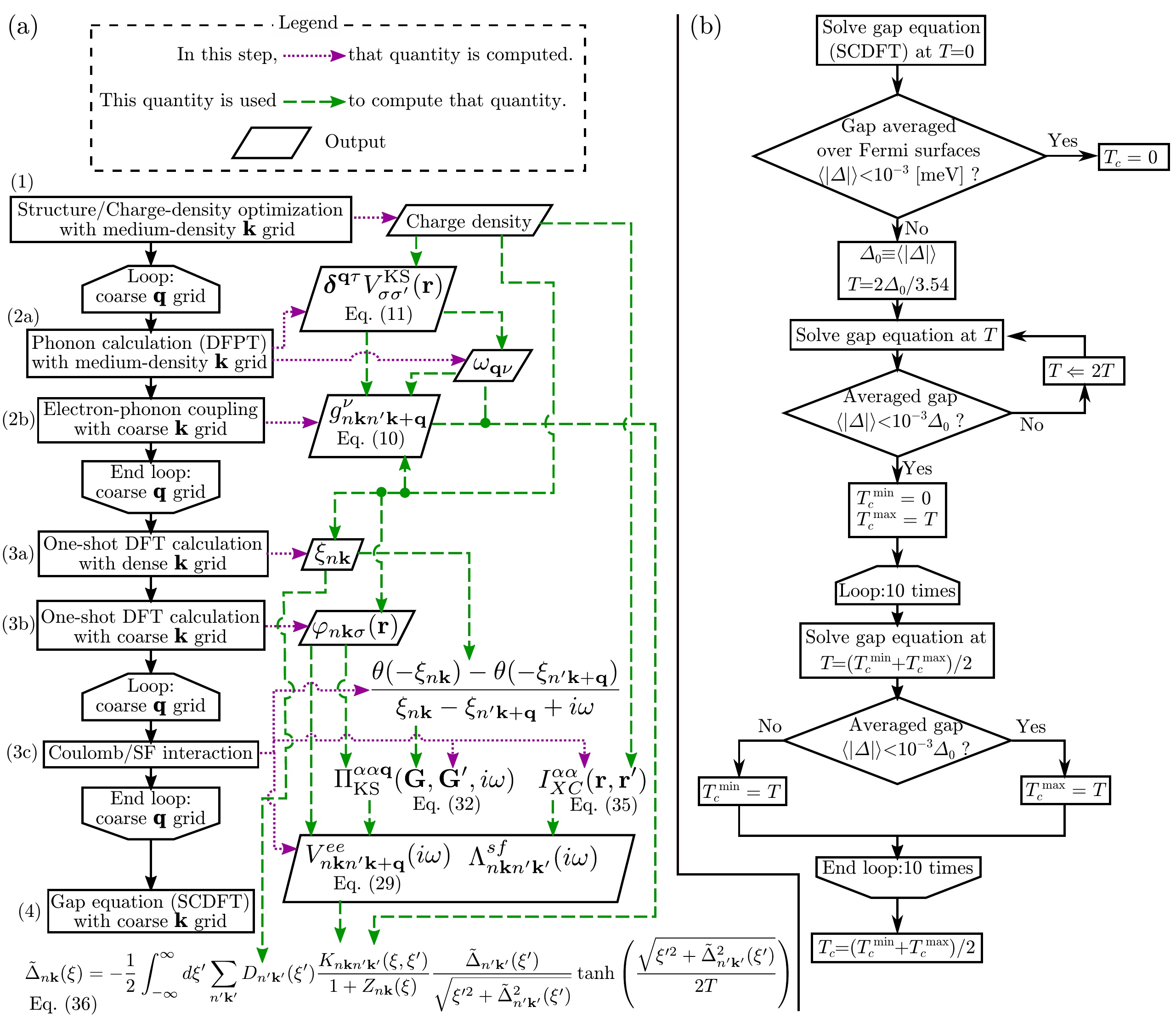}
    \caption{\label{fig_flow}
      (a) The flow chart to perform the SCDFT calculation for each material in this study.
      The indices of steps are the same as those mentioned in Sec.~\ref{subsec_overall}.
      (b) The flow chart of the bisection method used to compute $T_c$.}
  \end{center}
\end{figure*}
In this section, we will explain the practical procedure to perform the calculations
explained in the previous section.

\subsection{Evaluation of exchange integrals with Fourier transformation}

The exchange integrals with Coulomb interaction part $V_{n {\bf k} n' {\bf k}'}^{ee}$
 of Eq.~(\ref{eq_vee})
and the SF part $\Lambda_{n {\bf k} n' {\bf k}'}^{sf}$ of Eq.~(\ref{eq_lambdasf})
can be computed efficiently using the Fourier transformation as follows:
First $\rho^{(\alpha)}({\bf r})$ in Eqs. (\ref{eq_rho0}) and (\ref{eq_rhox})-(\ref{eq_rhoz})
has the periodicity of the lattice vector ${\bf R}$ together with the
phase factor from the Bloch theorem as
\begin{align}
  \rho_{n{\bf k}n'{\bf k}'}^{(\alpha)}({\bf r}+{\bf R})
  = e^{i({\bf k}'-{\bf k})\cdot{\bf R}} \rho_{n{\bf k}n'{\bf k}'}^{(\alpha)}({\bf r}).
\end{align}
Therefore $\rho^{(\alpha)}({\bf r})$ can be expanded with the Fourier components
of the reciprocal lattice vectors ${\bf G}$ as
\begin{align}
  \rho_{n{\bf k}n'{\bf k}+{\bf q}}^{(\alpha)}({\bf r})
  = \sum_{\bf G} e^{i ({\bf q} + {\bf G}) \cdot {\bf r}}
  {\tilde \rho}^{(\alpha)}_{n{\bf k}n'{\bf k}+{\bf q}}({\bf G}),
\end{align}
where ${\tilde \rho}^{(\alpha)}({\bf G})$ is defined by the Fourier transformation
of $\rho^{(\alpha)}({\bf r})$ as follows:
\begin{align}
  {\tilde \rho}^{(\alpha)}_{n{\bf k}n'{\bf k}+{\bf q}}({\bf G}) \equiv
  \frac{1}{v_{\rm uc}}\int_{\rm uc} d^3 r e^{-i ({\bf q} + {\bf G}) \cdot {\bf r}}
  \rho_{n{\bf k}n'{\bf k}+{\bf q}}^{(\alpha)}({\bf r}).
\end{align}
Subsequently, the exchange integral of Eq.~(\ref{eq_vee}) is rewritten as
\begin{align}
  V_{n {\bf k} n' {\bf k}+{\bf q}}^{ee} (i\omega)
  &= \sum_{{\bf G} {\bf G}'}
  V_{RPA}^{\bf q}({\bf G}, {\bf G}', i\omega)
  \nonumber \\
  &\hspace{1cm} \times
            {\tilde \rho}^{(0)}_{n {\bf k} n' {\bf k}+{\bf q}}({\bf G})
            {\tilde \rho}^{(0)*}_{n {\bf k} n' {\bf k}+{\bf q}}({\bf G}'),
\end{align}
where $V_{RPA}^{\bf q}$ is the
Fourier component of the screened Coulomb interaction as follows:
\begin{align}
  &V_{RPA}^{\bf q}({\bf G}, {\bf G}', i\omega)
  \nonumber \\
  &\equiv \iint d^3r d^3r' e^{i({\bf q}+{\bf G})\cdot{\bf r}}
  e^{-i({\bf q}+{\bf G}')\cdot{\bf r}'}
  V_{RPA} ({\bf r}, {\bf r}', i\omega).
  \label{eq_fourier_vrpa}
\end{align}
However, the values of $V_{RPA} ({\bf r}, {\bf r}’, i\omega)$ need not be found,
since we show that
$V_{RPA}^{\bf q}({\bf G}, {\bf G}', i\omega)$
at each ${\bf q}$ can be computed separately using the Bloch theorem as shown below.
By substituting $V_{RPA} ({\bf r}, {\bf r}', i\omega)$ of Eq.~(\ref{eq_vrpa})
into Eq.~(\ref{eq_fourier_vrpa}), we obtain 
\begin{align}
  &V_{RPA}^{\bf q}({\bf G}, {\bf G}', i\omega)
  = \frac{4\pi \delta_{{\bf G}{\bf G}'}}{|{\bf q}+{\bf G}|^2}
  \nonumber \\
  &+\sum_{{\bf G}_1} V_{RPA}^{\bf q}({\bf G}, {\bf G}_1, i\omega)
  \Pi^{00{\bf q}}_{\rm KS}({\bf G}_1, {\bf G}', i\omega)
  \frac{4\pi}{|{\bf q}+{\bf G}'|^2}
  \nonumber \\
  &=\left[\frac{|{\bf q}+{\bf G}'|^2\delta_{{\bf G}{\bf G}'}}{4\pi}
    - \Pi^{00{\bf q}}_{\rm KS}({\bf G}, {\bf G}', i\omega)\right]^{-1},
  \label{eq_fourier_vrpa_dyson}
\end{align}
where $\Pi^{\alpha\alpha{\bf q}}_{\rm KS}({\bf G}, {\bf G}', i\omega)$
is the Fourier component of the susceptibilities of the Kohn-Sham system
of Eq.~(\ref{eq_piks}) given by
\begin{align}
  &\Pi^{\alpha\alpha{\bf q}}_{\rm KS}({\bf G}, {\bf G}', i\omega)
  \nonumber \\
  &\equiv \iint d^3r d^3r' e^{i({\bf q}+{\bf G})\cdot{\bf r}}
  e^{-i({\bf q}+{\bf G}')\cdot{\bf r}'}
  \Pi^{\alpha\alpha}_{\rm KS} ({\bf r}, {\bf r}', i\omega)  
  \nonumber \\
  &= \sum_{{\bf k} n n'} 
  \frac{\theta(-\xi_{n {\bf k}})
    - \theta(-\xi_{n' {\bf k}+{\bf q}})}
       {\xi_{n {\bf k}}-\xi_{n' {\bf k}+{\bf q}}+i\omega}
  {\tilde \rho}^{(\alpha)}_{n {\bf k} n' {\bf k}+{\bf q}}({\bf G})
  {\tilde \rho}^{(\alpha)*}_{n {\bf k} n' {\bf k}+{\bf q}}({\bf G}').
  \label{eq_fourier_piks}
\end{align}
In the derivation of Eq.~(\ref{eq_fourier_vrpa_dyson}), we used
the periodicity of each term with respect to the lattice vector.
The factor $[\theta(-\xi_{n {\bf k}}) - \theta(-\xi_{n' {\bf k}+{\bf q}})]/(\xi_{n {\bf k}}-\xi_{n' {\bf k}+{\bf q}}+i\omega)$
in the susceptibilities
varies rapidly in the vicinity of Fermi surfaces,
and we need a dense ${\bf k}$ grid to compute it accurately,
which may require a huge numerical cost.
Therefore, we use the reverse interpolation scheme explained in Sec.~III.C.1 of
Ref.~\onlinecite{PhysRevB.95.054506}.
In this scheme, we compute the explicitly energy-dependent factor
with a dense ${\bf k}$ grid
while we compute the other parts using a coarse ${\bf k}$ grid because
${\tilde \rho}^{(\alpha)}_{n {\bf k} n' {\bf k}+{\bf q}}({\bf G}) {\tilde \rho}^{(\alpha)*}_{n {\bf k} n' {\bf k}+{\bf q}}({\bf G}')$ varies more smoothly
than the energy-dependent factor.
The SF term can be computed in the same manner at each ${\bf q}$ separately.
The Fourier component of the SF-mediated interaction of Eq.~(\ref{eq_hatlambdasf}) is
\begin{align}
  &\Lambda^{sf, \alpha \alpha}_{\bf q}({\bf G}, {\bf G}', i\omega)
  = - \sum_{{\bf G}_1, {\bf G}_2} I_{XC}^{\alpha \alpha{\bf q}} ({\bf G}, {\bf G}_1)
  \nonumber \\
  &\times
  \left[\left( \Pi_{\rm KS}^{\alpha \alpha{\bf q}}({\bf G}_1, {\bf G}_2)\right)^{-1}
    - I_{XC}^{\alpha \alpha{\bf q}}({\bf G}_1, {\bf G}_2)\right]^{-1}
  I_{XC}^{\alpha \alpha{\bf q}}({\bf G}_2, {\bf G}'),
  \label{eq_hatlambdasf_fourier}
\end{align}
where $I_{XC}^{\alpha \alpha{\bf q}}$ is the
Fourier component of the exchange correlation kernel of Eq.~(\ref{eq_xckernel}).
In this study, we employed the local density approximation (LDA) to describe this kernel as follows:
We approximate the exchange-correlation energy as
\begin{align}
  E_{XC}^{LDA} = \int d^3r \varepsilon_{XC}^{hom}(\rho({\bf r}), |{\bf m}({\bf r})|) \rho({\bf r}),
\end{align}
where $\rho({\bf r})$ is the electronic charge density and
$\varepsilon_{XC}^{hom}(\rho, m)$ is the exchange-correlation energy density of
the homogeneous electon gus whose charge and spin density are $\rho$ and $m$.
The exchange-correlation kernel in Eq.~(\ref{eq_xckernel}) becomes
\begin{align}
  &I_{XC}^{LDA, \alpha\alpha} ({\bf r},{\bf r}')
  \nonumber \\
  &\hspace{0.3cm}= \delta({\bf r}-{\bf r}') \rho({\bf r})
  \frac{\partial^2 \varepsilon(\rho({\bf r}),|{\bf m}|)}{\partial {|\bf m|}\partial {|\bf m|}}
  \left. \left(\frac{m_\alpha}{|{\bf m}|}\right)^2 \right|_{{\bf m}={\bf m}({\bf r})}.
\end{align}
Since we perform the non-magnetic calculation in this study,
we take the ${\bf m}({\bf r}) \rightarrow 0$ limit for this kernel.
Within LDA,
$I_{XC}^{\alpha \alpha{\bf q}}({\bf G}, {\bf G}')$ does not depend on ${\bf q}$.
This is equivalent to the adiabatic local density approximation (ALDA) \cite{PhysRevLett.55.2850} in time-dependent density functional theory \cite{PhysRevLett.52.997}.

\subsection{Auxiliary gap equation}

We have solved the gap equation (\ref{eq_gapeq}) using the auxiliary energy axis
\cite{PhysRevB.95.054506} to capture the rapid change of
the explicitly energy-dependent function in the vicinity of Fermi surfaces.
In this method, the gap function $\Delta_{n{\bf k}}$ depends also on the auxiliary energy;
the auxiliary gap function ${\tilde \Delta}_{n{\bf k}}(\xi)$
satisfies ${\tilde \Delta}_{n{\bf k}}(\xi_{n{\bf k}})=\Delta_{n{\bf k}}$.
Subsequently, the gap equation (\ref{eq_gapeq}) becomes
\begin{align}
  {\tilde \Delta}_{n {\bf k}} (\xi) = - \frac{1}{2} \int_{-\infty}^{\infty} d\xi'
  \sum_{n' {\bf k}'} D_{n' {\bf k}'}(\xi')
  \frac{K_{n {\bf k} n' {\bf k}'}(\xi, \xi') }
       {1 + Z_{n {\bf k}}(\xi)}
       \nonumber \\
       \times
       \frac{{\tilde \Delta}_{n' {\bf k}'}(\xi')}
            {\sqrt{\xi'^2 + {\tilde \Delta}^2_{n' {\bf k}'}(\xi')}}
       \tanh \left( \frac{\sqrt{\xi'^2 + {\tilde \Delta}^2_{n' {\bf k}'}(\xi')}}{2 T} \right),
       \label{eq_auxdelta}
\end{align}
where $D_{n {\bf k}}(\xi)$ is the $(n, {\bf k})$-resolved density of states.
In the same manner,
the electron-phonon and SF renormalization factors of Eqs. (\ref{eq_zep}) and (\ref{eq_zsf}) become
\begin{align}
  &Z_{n {\bf k}}^{ep}(\xi) =
  \frac{-1}{\tanh[\xi/(2T)]} \int_{-\infty}^{\infty} d\xi'
  \sum_{n' {\bf k}' \nu} D_{n' {\bf k}'}(\xi')
  |g_{n {\bf k} n' {\bf k}'}^{\nu}|^2
  \nonumber \\
  &\times[J(\xi, \xi', \omega_{{\bf k}'-{\bf k} \nu})
    + J(\xi, -\xi', \omega_{{\bf k}'-{\bf k} \nu})]
\end{align}
and
\begin{align}
  &Z_{n {\bf k}}^{sf}(\xi) =
  \frac{1}{\pi} \int_{-\infty}^{\infty} d\xi'
  \sum_{n'{\bf k}'} D_{n' {\bf k}'}(\xi') \int_0^\infty d\omega
  \nonumber \\
  &\times \frac{(|\xi|+|\xi_{n' {\bf k}'}|)^2 - \omega^2}
       {[ (|\xi|+|\xi_{n' {\bf k}'}|)^2+\omega^2 ]^2}
       \Lambda_{n {\bf k} n' {\bf k}'}^{sf}(i\omega),
       \label{eq_zsf_aux}
\end{align}
respectively,
where we employ the reverse interpolation method again;
the $(n, {\bf k})$-resolved density of states $D_{n {\bf k}}(\xi)$ is computed
with the dense ${\bf k}$ grid, while the other parts are computed on
the coarse ${\bf k}$ grid; finally, we combine the parts yielded.

\subsection{Frequency integral}

The integration in Eq.~(\ref{eq_kee}) involves
the frequency $\omega$ spanning [0,$\infty$].
Therefore, to perform the integration numerically,
we change the variable as follows:
\begin{align}
  \omega = (|\xi|+|\xi'|)\frac{1+x}{1-x}.
  \label{eq_omegatrans}
\end{align}
Then Eq.~(\ref{eq_kee}) becomes
\begin{align}
    K_{n {\bf k} n' {\bf k}'}^{ee}(\xi,\xi') =
  \frac{2}{\pi}\int_{-1}^{1} dx
  \frac{1}{1+x^2}
  \nonumber \\
  \times V_{n {\bf k} n' {\bf k}'}^{ee} \left(i (|\xi|+|\xi'|)\frac{1+x}{1-x}\right).
\end{align}
When $|\xi|+|\xi'|=0$, this integration becomes $V_{n {\bf k} n' {\bf k}'}^{ee} (0)$.
To obtain $V_{n {\bf k} n' {\bf k}'}^{ee}(i\omega)$ at an arbitrary $\omega$,
we first compute $V_{n {\bf k} n' {\bf k}'}^{ee}(i\omega)$ at
discrete non-uniform points $\omega = 0, \omega_1, \omega_2, \cdots, \infty$ and
interpolate them.
Using the same transformation as in Eq.~(\ref{eq_omegatrans}),
the frequency integral in the SF renormalization term (\ref{eq_zsf_aux})
is performed as follows:
\begin{align}
  Z_{n {\bf k}}^{sf}(\xi) &=
  \frac{1}{\pi} \int_{-\infty}^{\infty} d\xi'
  \sum_{n'{\bf k}'} D_{n' {\bf k}'}(\xi')
  \nonumber \\
  &\int_{-1}^{1} dx\frac{-2x}
            {(|\xi|+|\xi_{n' {\bf k}'}|)(1+x^2)^2}
            \nonumber \\
            &\times
            \Lambda_{n {\bf k} n' {\bf k}'}^{sf}
            \left(i (|\xi|+|\xi'|)\frac{1+x}{1-x}\right).
\end{align}
The numerical integration with respect to the variable $x$ ranging [-1,1]
can be performed using the Gauss quadrature.

\subsection{Overall procedure} \label{subsec_overall}

Figure~\ref{fig_flow} (a) shows the calculation flow.
We employ the following three different wavenumber grids
to efficiently perform the Brillouin-zone integrals:
\begin{description}
\item[coarse grid] To reduce the computational cost, we use a coarse grid for the wavenumber ${\bf q}$ of phonons and susceptibilities.
  The grid is shifted by half of its step to avoid the singularity at the $\Gamma$ point.
  This grid is also used for solving the gap equation.
\item[medium-density grid] The atomic structure and the charge density are optimized with the self-consistent field calculation using ${\bf k}$ grid denser than the coarse grid.
  This ${\bf k}$ grid is used to prepare electronic states in the DFPT calculation.
\item[dense grid] To treat the explicitly energy-dependent factor in the calculations
  of susceptibilities and the ($n, {\bf k}$)-dependent density of states
  in the gap equation (\ref{eq_auxdelta}), a dense ${\bf k}$ grid is employed.
\end{description}
The overall calculations are performed as follows:
\begin{enumerate}
\item First, we optimize the atomic structure and the charge density using the standard density functional calculation with the medium-density ${\bf k}$ grid.
  The following calculation is performed on this optimized atomic structure and with the charge density.
\item We compute the electron-phonon interaction and the frequency of phonons,
  whose wavenumber ${\bf q}$ is on the half-grid shifted coarse grid.
  This step is further split into the following two sub-steps:
  \begin{enumerate}
  \item The phonon calculation based on DFPT is performed;
    the electronic states used in this calculation have a wavenumber ${\bf k}$ on the medium-density grid.
  \item Electron-phonon vertex of Eq.~(\ref{eq_gep}) between the Kohn-Sham orbitals
    $(n, {\bf k})$ and $(n',{\bf k}+{\bf q})$ is computed,
    where the wavenumber ${\bf k}$ is on the coarse grid.
  \end{enumerate}
\item Next, we compute the exchange integrals of screened Coulomb and SF-mediated interaction
  whose transitional momentum ${\bf q}$ is on the half-grid shifted coarse grid.
  This step is split into the following three sub-steps:
  \begin{enumerate}
  \item One-shot DFT calculation on the dense ${\bf k}$ grid is performed.
    The resulting energy dispersion $\xi_{n{\bf k}}$ is later used to compute the explicitly energy-dependent term in the susceptibilities in Eq. (\ref{eq_fourier_piks}).
  \item One-shot DFT calculation on the coarse ${\bf k}$ grid with and without a half-grid shift
    is performed.
    These two grids are connected by the transitional momentum ${\bf q}$ on the coarse a half-grid shifted grid.
    The resulting Kohn-Sham orbitals will be used to compute
    $\rho^{(\alpha)}({\bf r})$ of Eqs. (\ref{eq_rho0}) and (\ref{eq_rhox})-(\ref{eq_rhoz}).
  \item The exchange integrals of screened Coulomb and SF-mediated interaction
    between the Kohn-Sham orbitals
    $(n, {\bf k})$ and $(n',{\bf k}+{\bf q})$ are computed,
    where the wavenumber ${\bf k}$ is on the coarse grid.
  \end{enumerate}
\item The gap equation within SCDFT is solved on the coarse ${\bf k}$ grid at each temperature.
  Then $T_c$ is obtained as a minimum temperature where all $\Delta_{n{\bf k}}(\xi)$ vanish.
\end{enumerate}
The transition temperature $T_c$ is found using the bisection method explained
in Fig.~\ref{fig_flow} (b).
While the initial lower limit of $T_c$ is set to zero,
the initial upper limit is set to the $T_c$ estimated by the Bardeen-Cooper-Schrieffer theory
\cite{PhysRev.108.1175,schrieffer1983theory}
($2\Delta_0/3.54$, where $\Delta_0$ is the superconducting gap
averaged over Fermi surfaces at zero kelvin).
If there is a finite gap even at this upper limit,
although it rarely occurs, we double the initial upper limit;
then we repeat the bisection step ten times and find $T_c$.

\section{Result} \label{sec_result}

\begin{table}[!tb]
  \begin{center}
    \caption{\label{tbl_condition}
      Structure, cutoff for the plane wave for wave functions,
      ${\bf q}$-grid for phonon, and the error of the lattice constant
      $\langle \Delta a/a_{\rm exp}\rangle\equiv (V_{\rm calc}/V_{\rm exp})^{1/3} - 1$.
      $V_{\rm calc}$ and $V_{\rm exp}$ are the calculated and experimental
      unit-cell volume, respectively.
    }
    \begin{tabular}{ccccc}
      & structure & cutoff [Ry] & coarse grid & $\langle \Delta a/a_{\rm exp}\rangle$ [\%]\\
      \hline
      Be & hcp & 65 & $10\times10\times5$ & -0.74 \\
      \hline
      Na & bcc & 90 & $6\times6\times6$ & -0.60 \\
      Mg & hcp & 65 & $7\times7\times4$ & -0.77 \\
      Al & fcc & 65 & $8\times8\times8$ & -0.90 \\
      \hline
      K & bcc & 120 & $5\times5\times5$ & 0.61 \\
      Ca & fcc & 120 & $6\times6\times6$ & -1.20 \\
      Sc & hcp & 45 & $7\times7\times4$ & -0.49 \\
      Ti & hcp & 50 & $7\times7\times4$ & -0.28 \\
      V & bcc & 100 & $9\times9\times9$ & -1.19 \\
      Cu & fcc & 90 & $9\times9\times9$ & 0.42 \\
      Zn & hcp & 90 & $8\times8\times4$ & -0.02 \\
      Ga & $\alpha$-Ga & 150 & $5\times5\times4$ & 1.35 \\
      \hline
      Rb & bcc & 30 & $5\times5\times5$ & 0.58 \\
      Sr & fcc & 30 & $5\times5\times5$ & -1.09 \\
      Y & hcp & 40 & $6\times6\times3$ & 0.94 \\
      Zr & hcp & 50 & $7\times7\times4$ & 0.09 \\
      Nb & bcc & 90 & $8\times8\times8$ & 0.53 \\
      Mo & bcc & 35 & $8\times8\times8$ & 0.44 \\
      Tc & hcp & 30 & $8\times8\times4$ & 0.04 \\
      Ru & hcp & 35 & $8\times8\times4$ & -0.62 \\
      Rh & fcc & 35 & $8\times8\times8$ & 0.73 \\
      Pd & fcc & 45 & $8\times8\times8$ & 1.18 \\
      Ag & fcc & 50 & $8\times8\times8$ & 1.42 \\
      Cd & hcp & 45 & $7\times7\times3$ & 2.32 \\
      In & bct & 65 & $7\times7\times7$ & 1.84 \\
      Sn & $\beta$-Sn & 50 & $7\times7\times7$ & 1.72 \\
      \hline
      Cs & bcc & 75 & $4\times4\times4$ & 1.01 \\
      Ba & bcc & 30 & $5\times5\times5$ & -0.19 \\
      La & hcp & 120 & $6\times6\times3$ & 0.60 \\
      Hf & hcp & 55 & $7\times7\times4$ & 0.01 \\
      Ta & bcc & 50 & $8\times8\times8$ & 0.79 \\
      W & bcc & 50 & $8\times8\times8$ & 0.62 \\
      Re & hcp & 60 & $8\times8\times4$ & 0.39 \\
      Os & hcp & 55 & $8\times8\times4$ & 1.00 \\
      Ir & fcc & 40 & $8\times8\times8$ & 1.29 \\
      Pt & fcc & 60 & $8\times8\times8$ & 1.15 \\
      Au & fcc & 45 & $8\times8\times8$ & 1.15 \\
      Hg & trigonal & 50 & $6\times6\times6$ & 6.81 \\
      Tl & hcp & 55 & $6\times6\times3$ & 3.10 \\
      Pb & fcc & 35 & $6\times6\times6$ & 1.77 \\
      \hline
    \end{tabular}
  \end{center}
\end{table}
In this section we will first explain the numerical condition of this study,
then show the result of the benchmark.

The numerical condition is as follows:
We use the DFT code {\sc Quantum ESPRESSO} \cite{0953-8984-29-46-465901}
which employs plane waves and pseudopotentials.
Perdew-Burke-Ernzerhof's density functional \cite{PhysRevLett.78.1396}
based on generalized gradient approximation (GGA) is used.
We use the optimized norm-conserving pseudopotential \cite{PhysRevB.88.085117} library
provided by Schlipf-Gygi (SG15) \cite{SCHLIPF201536, Prandini2018}.
The energy cutoff for the wave functions of each element is specified using the criteria and
the convergence profiles in the Standard Solid State Pseudopotentials \cite{Prandini2018}.
We are using the optimized tetrahedron method \cite{PhysRevB.89.094515} to perform the Brillouin-zone integration.
The number of grids along each reciprocal lattice vector is proportional to the length of that vector.
Table~\ref{tbl_condition} presents the above explained conditions for each element.
The SCDFT calculation is performed using the 
{\sc Superconducting Toolkit} \cite{SuperconductingToolkit} software package.
We have set the medium-density grid twice the density of the coarse grid and set the dense grid twice the density of the medium-density grid;
for example, $8^3$, $16^3$, and $32^3$ grids were used for the coarse, medium-density,
and dense grids, respectively; all for Al.
The minimum scale and the number of points of the non-uniform auxiliary energy grid used in the gap equation (\ref{eq_auxdelta}) were set to $10^{-7}$ Ry and 100, respectively.
In the calculation of the magnetic exchange-correlation kernel of Eq.~(\ref{eq_xckernel}),
we ignore the gradient correction.
For stabilizing the phonon calculation, the lattice constants and the internal atomic coordinates are optimized;
the deviation between the optimized and the experimental lattice constants are listed in Tbl.~\ref{tbl_condition}.
To compute the susceptibilities in Eq.~(\ref{eq_fourier_piks}) and solve the gap equation (\ref{eq_auxdelta}), we have included $40 \times N_{\rm atom}$ ($20 \times N_{\rm atom}$) empty bands for the calculation with (without) SOI, where $N_{\rm atom}$ is the number of atoms per unit cell.

\begin{turnpage}
  \begin{figure*}[tb!]
    \begin{center}
      \includegraphics[width=23.0cm]{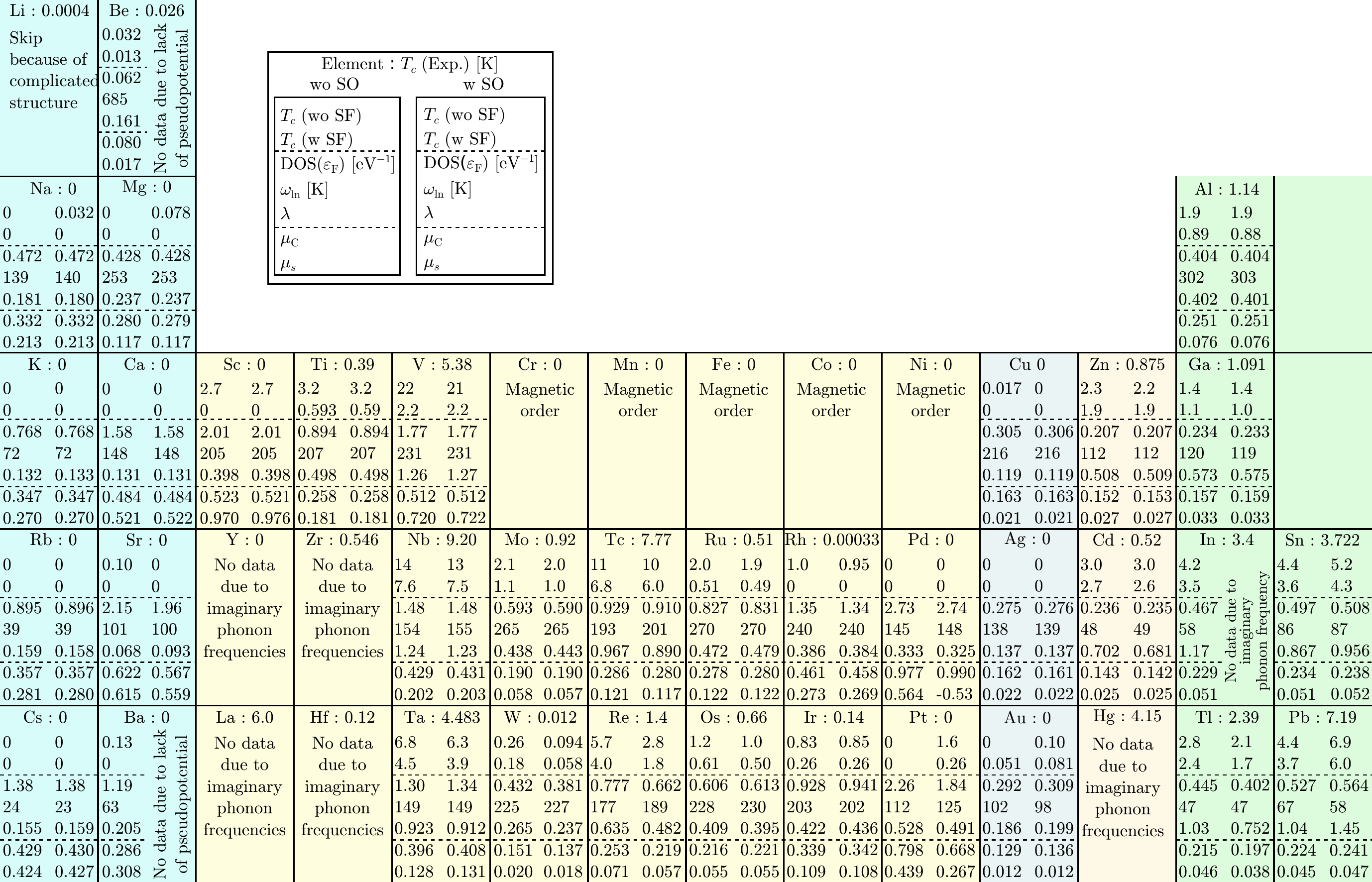}
      \caption{\label{fig_periodic}
        The experimental $T_c$ ($T_c^{\rm exp}$) \cite{HAMLIN201559} and
        the theoretical $T_c$ computed with and without SOI/SF.
        In addition, we show
        the density of states (DOS)
        divided by the number of atoms,
        the averaged phonon frequencies $\omega_{\rm ln}$,
        the Fr\"ohlich's mass-enhancement parameter $\lambda$,
        the averaged Coulomb interaction $\mu_{\rm C}$ in Eq.~(\ref{eq_mu_c}),
        and
        the averaged SF term $\mu_s$ in Eq.~(\ref{eq_mu_s}).
        Cyan, green, red, blue, and yellow cells indicate
        the alkaline (earth) metals, the sp-orbital metals,
        the group 12 metals, the noble metals, and the other transition metals, respectively.
      }
    \end{center}
  \end{figure*}
\end{turnpage}
\begin{figure*}[!tb]
  \begin{center}
    \includegraphics[width=16cm]{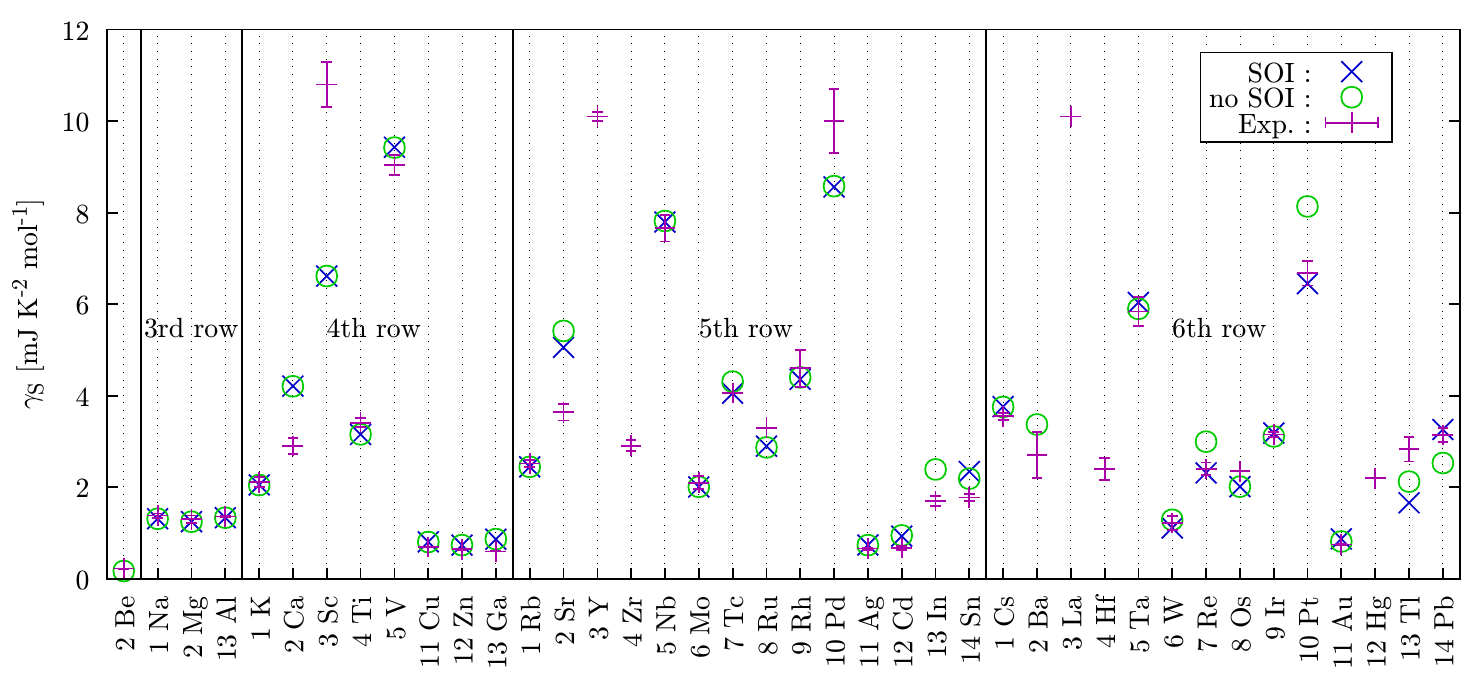}
    \caption{\label{fig_sommerfeld}
      Theoretical and experimental \cite{GSCHNEIDNER1964275} Sommerfeld coefficient $\gamma_{\rm S}$.
      The horizontal axis is the atomic symbol together with the group of the periodic table.
      Blue crosses and green circles indicate the Sommerfeld coefficient computed with and without SOI
      while magenta ``+'' with an error bar indicates the experimental value of the Sommerfeld coefficient.}
  \end{center}
\end{figure*}
\begin{figure*}[!tb]
  \begin{center}
    \includegraphics[width=16cm]{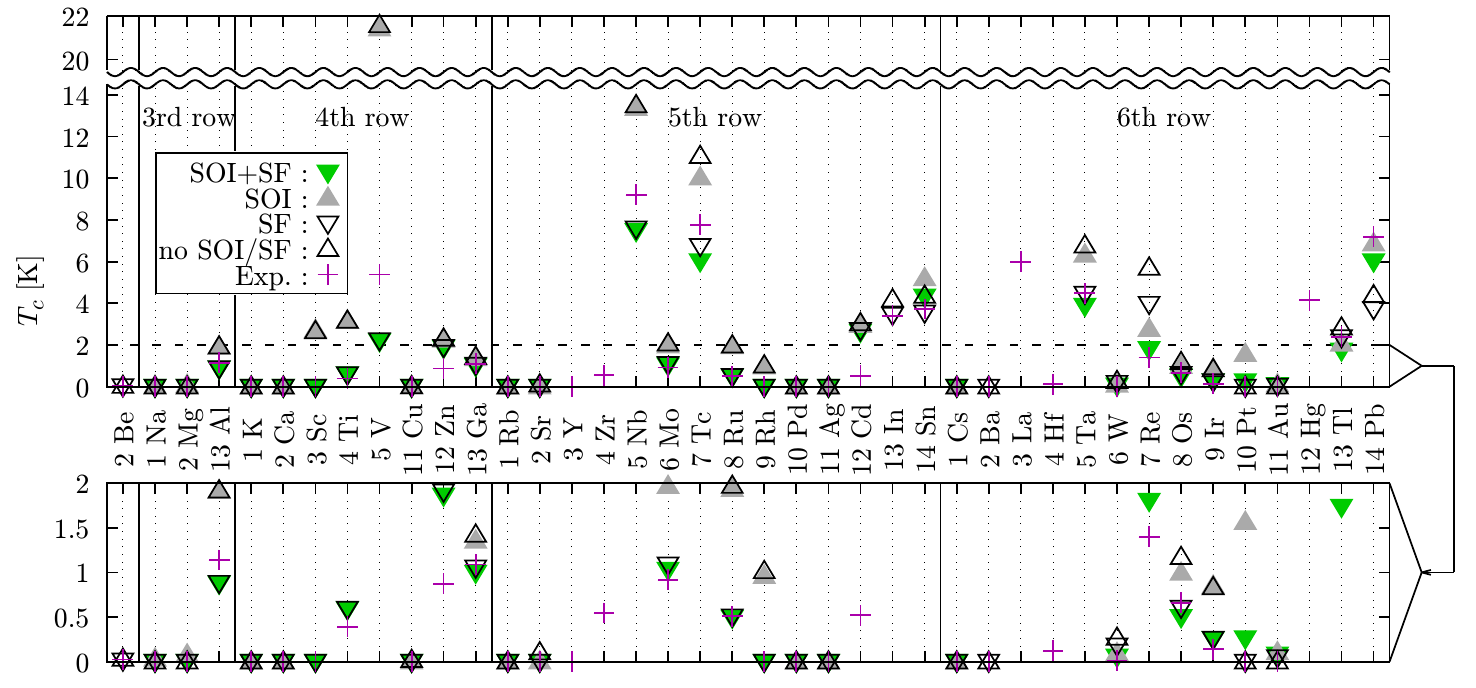}
    \caption{\label{fig_t_c}
      Theoretical and experimental $T_c$s.
      The horizontal axis is the atomic symbol together with the group of the periodic table.
      Downward (upward) triangles indicate the $T_c$s computed with (without) SF.
      Filled (empty) triangles indicate the $T_c$s computed with (without) SOI.
      ``+'' indicates the experimental value of $T_c$.
      The plot which ranges from zero to two Kelvin in the upper panel is
      magnified into the bottom panel.}
  \end{center}
\end{figure*}
Next, we move onto the results.
Figure~\ref{fig_periodic} shows 
the experimental $T_c$ ($T_c^{\rm exp}$) \cite{HAMLIN201559},
the theoretical $T_c$ computed with and without SOI/SF in a periodic-table form.
Also, to examine the effects of
the electron-phonon interaction, the screened Coulomb repulsion,
and the spin-fluctuation, 
we are showing the following quantities in the same figure:
The density of states (DOS) at the Fermi level
divided by the number of atoms affects the strength of the mean-field.
The Fr\"ohlich's mass-enhancement parameter
\begin{align}
  \lambda = \sum_{{\bf q} \nu} \lambda_{{\bf q}\nu},
  \label{eq_frohlich}
\end{align}
and the averaged phonon frequencies
\begin{align}
  \omega_{\rm ln} =\exp\left[ \frac{1}{\lambda}
    \sum_{{\bf q} \nu} \lambda_{{\bf q}\nu} \ln(\omega_{{\bf q}\nu})\right],
  \label{eq_omegaln}
\end{align}
appear in the conventional McMillan formula \cite{PhysRev.167.331,Dynes1972615}
\begin{align}
  T_c = \frac{\omega_{\rm ln}}{1.2}\exp\left[
    \frac{-1.04(1+\lambda)}{\lambda-\mu^*(1+0.62\lambda)}
    \right]
\end{align}
which has been used to estimate $T_c$ semi-empirically
with an adjustable parameter $\mu^*$.
The $({\bf q}, \nu)$-dependent mass-enhancement parameter $\lambda_{{\bf q} \nu}$
is computed as follows:
\begin{align}
  \lambda_{{\bf q} \nu} = \frac{2}{D(0)\omega_{{\bf q}\nu}}
  \sum_{{\bf k} n n'} |g_{n {\bf k} n' {\bf k}+{\bf q}}^{\nu}|^2
  \delta(\xi_{n {\bf k}}) \delta(\xi_{n' {\bf k}+{\bf q}}),
  \label{eq_lambda_q_nu}
\end{align}
where $D(0)$ is the density of states at the Fermi level.
We are calculating here the Brillouin-zone integral, including two delta functions, using the dense ${\bf k}$ grid together with the optimized tetrahedron method \cite{PhysRevB.89.094515}.
Because of the double delta function $\delta(\xi_{n {\bf k}}) \delta(\xi_{n' {\bf k}+{\bf q}})$,
this summation involves the electron-phonon vertices only between the electronic states
at the Fermi level.
Therefore, the Fr\"ohlich's mass-enhancement parameter $\lambda$ in Eq.~(\ref{eq_frohlich}) indicates
the retarded phonon-mediated interaction ($2|g|^2/\omega$) averaged over Fermi surfaces
times the density of state at the Fermi level.
Similarly, $\omega_{\rm ln}$ indicates the typical frequency of phonons which couples largely
with the electronic states at the Fermi level.
Therefore, $\lambda$ and $\omega_{\rm ln}$ closely relate to the electron-phonon contribution to $T_c$.
In an analogous fashion to $\lambda$, we are showing parameters for
the Coulomb repulsion and SF as
\begin{align}
  \mu_{\rm C} = \frac{1}{D(0)}
  \sum_{{\bf k} {\bf k}' n n'} K_{n {\bf k} n' {\bf k}'}^{ee}
  \delta(\xi_{n {\bf k}})\delta(\xi_{n' {\bf k}'}),
  \label{eq_mu_c}
\end{align}
and
\begin{align}
  \mu_{s} = \frac{1}{D(0)}
  \sum_{{\bf k} {\bf k}' n n'} K_{n {\bf k} n' {\bf k}'}^{sf}
  \delta(\xi_{n {\bf k}})\delta(\xi_{n' {\bf k}'}),
  \label{eq_mu_s}
\end{align}
respectively.
These parameters are the Coulomb [Eq.(\ref{eq_kee})] and SF [Eq.(\ref{eq_ksf})] kernels
averaged over Fermi surfaces times the density of states at the Fermi level.

We note that in Fig.~(\ref{fig_periodic}), some results are absent because of the following reason:
Li has a vast unit cell at a low temperature, and it is computationally demanding.
Cr, Mn, Fe, Co, and Ni show a magnetic order at the low temperature.
Therefore,
the formalism of the spin-fluctuation (\ref{eq_hatlambdasf}) used in this study breaks down;
the matrix
\begin{align}
  \delta_{{\bf G} {\bf G}'} - \sum_{{\bf G}_1}
  \Pi_{\rm KS}^{\alpha \alpha{\bf q}}({\bf G}, {\bf G}_1)
  I_{XC}^{\alpha \alpha{\bf q}}({\bf G}_1, {\bf G}')
  \label{eq_stoner_criterion}
\end{align}
does not become
positive definite (the Stoner's criterion) for those materials.
For Be and Ba, there is no pseudopotential together with SOI in the pseudopotential library
used in this study.
Since we are trying to unify the condition of the calculation for all elements,
we leave the result of these two materials together with SOI blank.
For Y, Zr, In (with SOI), La, Hf, and Hg,
we have obtained imaginary phonon frequencies because of an artificial long-range structure instability.
For such cases,
we could have not continue the calculation because of the breakdown of the formulations of the electron-phonon kernel [Eq.~(\ref{eq_kep})] and renormalization term [Eq.~(\ref{eq_zep})].
Therefore, we leave the results for those cases blank.

\begin{figure}[!tb]
  \begin{center}
    \includegraphics[width=8cm]{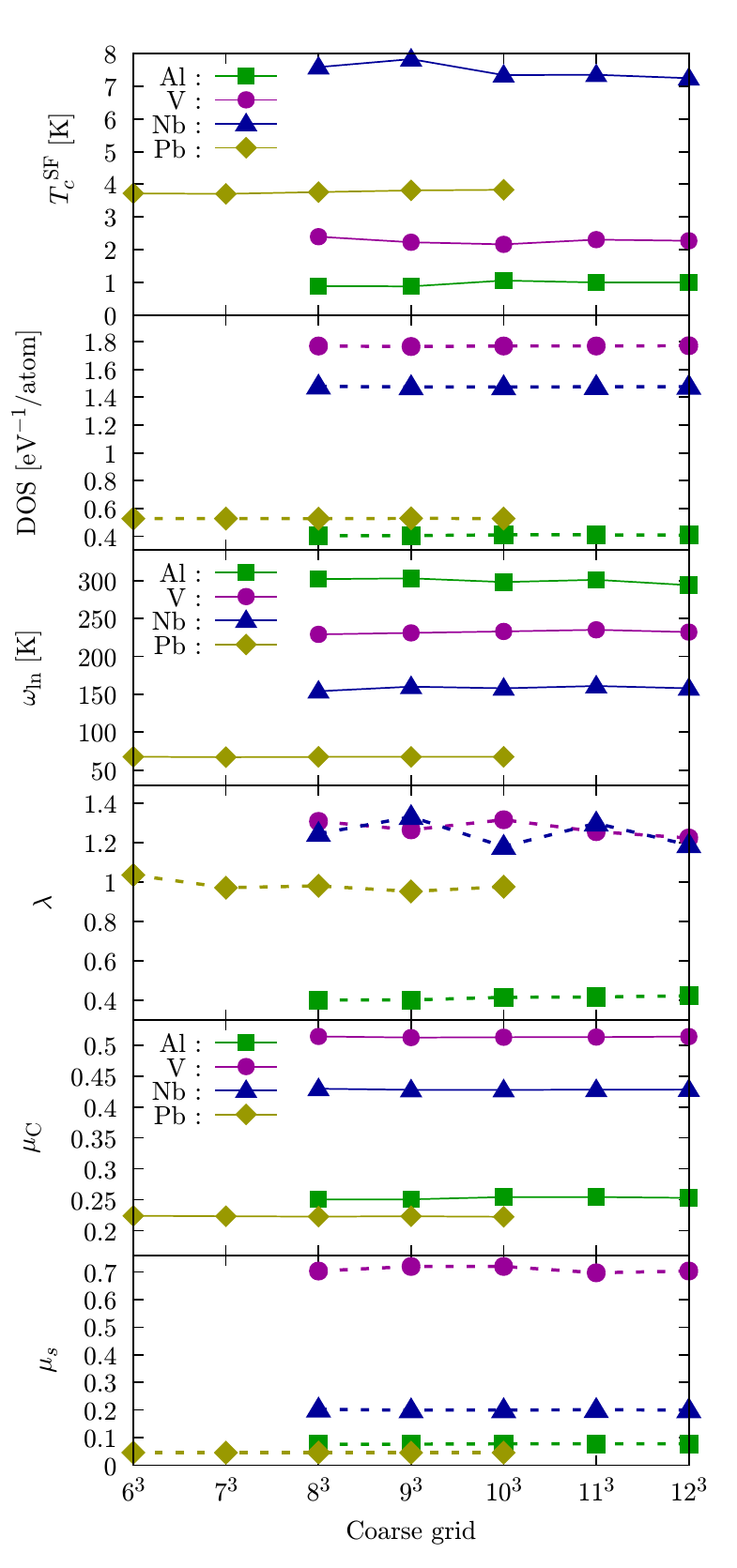}
    \caption{\label{fig_conv}
      The convergence of calculated $T_c$ (with SF, without SOI), the density of states at the Fermi level, the averaged phonon frequencies $\omega_{\rm ln}$, the Fr\"ohlich's mass-enhancement parameter $\lambda$, he averaged Coulomb interaction $\mu_{\rm C}$ in Eq.~(\ref{eq_mu_c}), and the averaged SF term $\mu_s$ in Eq.~(\ref{eq_mu_s}) with respect to the density of the wave-number grids. We pick up Al, V, Nb, and Pb.
      }
  \end{center}
\end{figure}

\begin{figure}[!tb]
  \begin{center}
    \includegraphics[width=8cm]{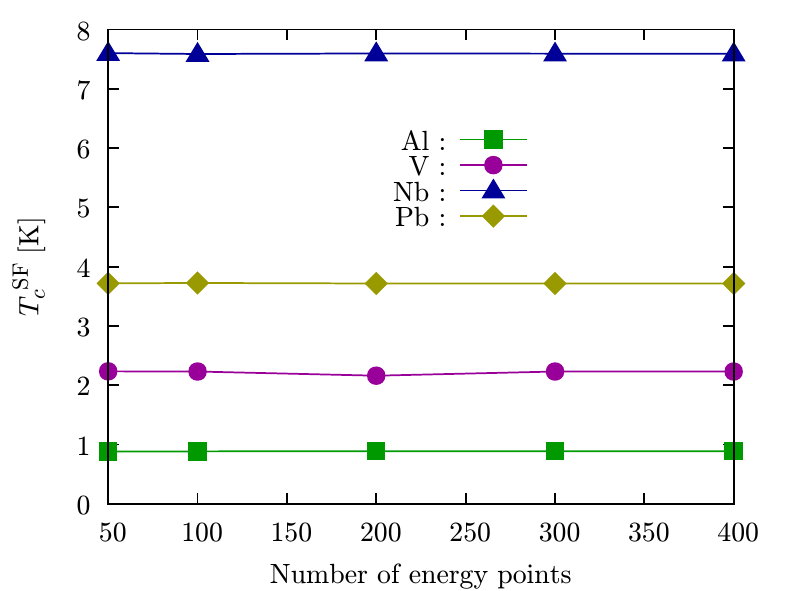}
    \caption{\label{fig_conv_ene}
      The convergence of the $T_c$ (with SF) with respect to the number of points for the energy integral in Eq.~(\ref{eq_auxdelta}) for Al, V, Nb, and Pb.
      }
  \end{center}
\end{figure}

We have checked the convergences of the numerical results with respect to the density of ${\bf k}$, ${\bf q}$, and the auxiliary energy grids.
For this purpose, we have selected up the typical four materials, namely Al, V, Nb, and Pb.
Figure~\ref{fig_conv} shows the convergence of calculated $T_c$ (with SF, without SOI), the density of states at the Fermi level, the averaged phonon frequencies $\omega_{\rm ln}$, the Fr\"ohlich's mass-enhancement parameter $\lambda$, he averaged Coulomb interaction $\mu_{\rm C}$ in Eq.~(\ref{eq_mu_c}), and the averaged SF term $\mu_s$ in Eq.~(\ref{eq_mu_s}) with respect to the density of the wave-number grids.
Also, in the convergence check, we have set the medium-density grid twice the density of the coarse grid and have set the dense grid twice the density of the medium-density grid.
Although the $\lambda$ of V and Nb oscillate within 4\% and 7\%, respectively, when we change the density of the coarse grid, other quantities, including $T_c$, are unchanged.
We are showing the convergence of $T_c$ (with SF, without SOI) with respect to the number of points for the energy integral in Eq.~(\ref{eq_auxdelta}) in Fig.~\ref{fig_conv_ene}.
This result shows the $T_c$s have converged at the numerical conditions described above.

We compare $\lambda$ in Eq.~(\ref{eq_frohlich}), $\omega_{\rm ln}$ in Eq.~(\ref{eq_omegaln}),
and $\mu_{\rm C}$ in Eq.~(\ref{eq_mu_c})
obtained in this study and earlier studies.
Table~\ref{tbl_compare_lambda} shows $\lambda$ and $\omega_{\rm ln}$ in this study and earlier studies.
Although there are small deviations due to the different exchange-correlation functional,
these quantities are consistent with those of earlier studies excepting the case for Pb without SOI;
the reported $\lambda$s of Pb without SOI are scattered and different from the experimental value ($\lambda_{\rm exp} = 1.55$) \cite{lambda_of_pb}.
We also confirm that we reproduced the averaged Coulomb interaction $\mu_{\rm C}$ in Eq.~(\ref{eq_mu_c})
in the earlier works for Al and Nb;
$\mu_{\rm C}$ is 0.251 and 0.429 for Al and Nb, respectively in this study
while those in the earlier studies are 0.236~\cite{PhysRevB.52.1425} and 0.488~\cite{PhysRevB.54.1419}.

\begin{table*}[!tb]
  \begin{center}
    \caption{\label{tbl_compare_lambda}
      Fr\"ohlich's mass-enhancement parameter $\lambda$ in Eq.~(\ref{eq_frohlich}) and
      the averaged phonon frequencies $\omega_{\rm ln}$ in Eq.~(\ref{eq_omegaln})
      computed in this study and earlier studies.
      Numerical conditions,
      the ${\bf q}$ grid of phonons,
      the ${\bf k}$ grid of electronic states for the calculation of phonon frequencies,
      the ${\bf k}$ grid and the smearing width for the integration in Eq.~(\ref{eq_lambda_q_nu}) are also written.
      In the Ref. \onlinecite{PhysRevB.53.R7575} only the number of ${\bf q}$ and ${\bf k}$ points in the irreducible Brillouin-zone (IBZ) are provided.
      In this study and one of the earlier studies~\cite{PhysRevB.54.16487}
      the tetrahedron method is employed for performing the Brillouin-zone integration in
      Eq.~(\ref{eq_lambda_q_nu}).
    }
    \begin{tabular}{ccccccccc}
      & $\lambda$ & $\omega_{\rm ln}$ [K] & ${\bf q}$ grid & ${\bf k}$ grid (phonon) &
      ${\bf k}$ grid ($\lambda$) & Smearing [eV] & Functional & Ref. \\
      \hline
      Al (without SOI)
      & 0.402 & 302 & $8\times8\times8$ & $16\times16\times16$ & $32\times32\times32$ &Tetrahedron& GGA & This work \\
      & 0.438 & - & 89 in IBZ & - & 1,300 in IBZ & 0.272 & LDA & \onlinecite{PhysRevB.53.R7575} \\
      & 0.417 & 314 & $8\times8\times8$ & $12\times12\times12$ & $32\times32\times32$ & 0.272 & LDA & \onlinecite{PhysRevLett.111.057006} \\
      & 0.44 & 270 & $8\times8\times8$ & $16\times16\times16$ & $32\times32\times32$ &Tetrahedron& LDA & \onlinecite{PhysRevB.54.16487} \\
      \hline
      V (without SOI) 
      & 1.26 & 231 & $9\times9\times9$ & $18\times18\times18$ & $36\times36\times36$ &Tetrahedron& GGA & This work \\
      & 1.19 & 245 & $8\times8\times8$ & $16\times16\times16$ & $32\times32\times32$ &Tetrahedron& LDA & \onlinecite{PhysRevB.54.16487} \\
      \hline
      Cu (without SOI)
      & 0.119 & 216 & $9\times9\times9$ & $18\times18\times18$ & $36\times36\times36$ &Tetrahedron& GGA & This work \\
      & 0.14 & 220 & $8\times8\times8$ & $16\times16\times16$ & $32\times32\times32$ &Tetrahedron& LDA & \onlinecite{PhysRevB.54.16487} \\
      \hline
      Nb (without SOI)
      & 1.24 & 154 & $8\times8\times8$ & $16\times16\times16$ & $32\times32\times32$ &Tetrahedron& GGA & This work \\
      & 1.26 & 185 & $8\times8\times8$ & $16\times16\times16$ & $32\times32\times32$ &Tetrahedron& LDA & \onlinecite{PhysRevB.54.16487} \\
      \hline
      Mo (without SOI)
      & 0.438 & 265 & $8\times8\times8$ & $16\times16\times16$ & $32\times32\times32$ &Tetrahedron& GGA & This work \\
      & 0.42 & 280 & $8\times8\times8$ & $16\times16\times16$ & $32\times32\times32$ &Tetrahedron& LDA & \onlinecite{PhysRevB.54.16487} \\
      \hline
      Pd (without SOI)
      & 0.333 & 145 & $8\times8\times8$ & $16\times16\times16$ & $32\times32\times32$ &Tetrahedron& GGA & This work \\
      & 0.35 & 180 & $8\times8\times8$ & $16\times16\times16$ & $32\times32\times32$ &Tetrahedron& LDA & \onlinecite{PhysRevB.54.16487} \\
      \hline
      Ta (without SOI)
      & 0.923 & 149 & $8\times8\times8$ & $16\times16\times16$ & $32\times32\times32$ &Tetrahedron& GGA & This work \\
      & 0.86 & 160 & $8\times8\times8$ & $16\times16\times16$ & $32\times32\times32$ &Tetrahedron& LDA & \onlinecite{PhysRevB.54.16487} \\
      \hline
      Tl (without SOI)
      & 1.03 & 47 & $6\times6\times3$ & $12\times12\times6$ & $24\times24\times12$ &Tetrahedron& GGA & This work \\
      & 1.0 & - & $8\times8\times8$ & $24\times24\times16$ & $36\times36\times24$ & 0.2 & LDA & \onlinecite{PhysRevB.81.174527} \\
      \hline
      Tl (with SOI) 
      & 0.752 & 47 & $6\times6\times3$ & $12\times12\times6$ & $24\times24\times12$ &Tetrahedron& GGA & This work \\
      & 0.87 & - & $8\times8\times8$ & $24\times24\times16$ & $36\times36\times24$ & 0.2 & LDA & \onlinecite{PhysRevB.81.174527} \\
      \hline
      Pb (without SOI)
      & 1.04 & 67 & $6\times6\times6$ & $12\times12\times12$ & $24\times24\times24$ &Tetrahedron& GGA & This work \\
      & 1.20 & - & 89 in IBZ   & -                    & 1,300 in IBZ   & 0.272 & LDA & \onlinecite{PhysRevB.53.R7575} \\
      & 1.68 & 65 & $8\times8\times8$ & $16\times16\times16$ & $32\times32\times32$ &Tetrahedron& LDA & \onlinecite{PhysRevB.54.16487} \\
      & 1.08 & - & $8\times8\times8$ & $16\times16\times16$ & $32\times32\times32$ & 0.2 & LDA & \onlinecite{PhysRevB.81.174527} \\
      & 1.24 & - & $8\times8\times8$ & $16\times16\times16$ & $40\times40\times40$ & 0.272 & LDA & \onlinecite{PhysRevB.87.024505} \\
      \hline
      Pb (with SOI)
      & 1.45 & 58 & $6\times6\times6$ & $12\times12\times12$ & $24\times24\times24$ &Tetrahedron& GGA & This work \\
      & 1.56 & - & $8\times8\times8$ & $16\times16\times16$ & $32\times32\times32$ & 0.2 & LDA & \onlinecite{PhysRevB.81.174527} \\
      \hline
    \end{tabular}
  \end{center}
\end{table*}

\begin{figure}[!tb]
  \begin{center}
    \includegraphics[width=8cm]{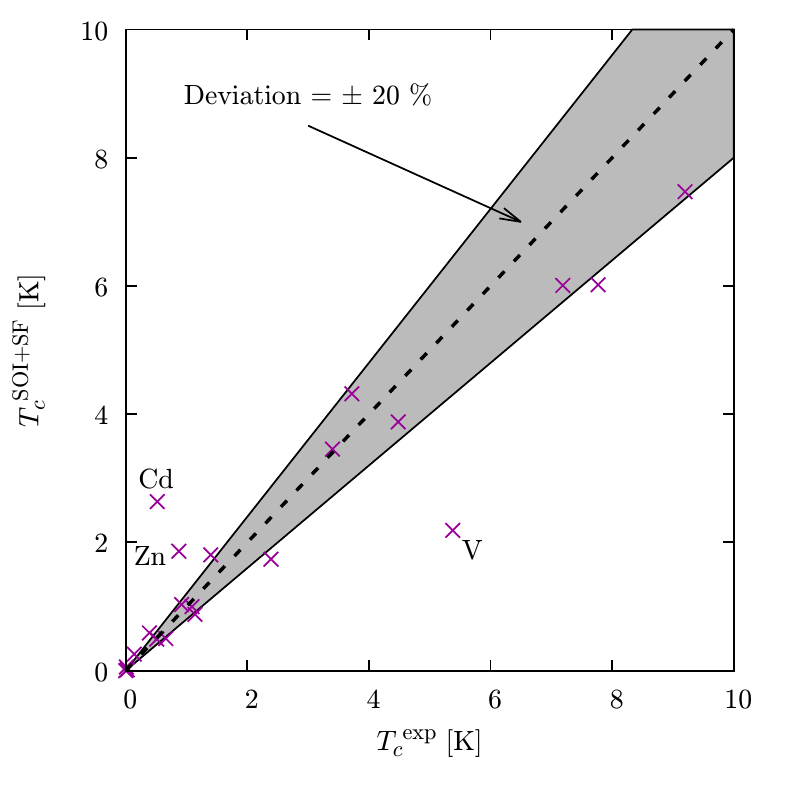}
    \caption{\label{fig_tc2}
      Experimental and computed $T_c$.
      For the theoretical value, only the result computed with SOI and SF is represented.
      The shaded region indicates that the deviation of $T_c$ ($T_c^{\rm SOI+SF}/T_c^{\rm exp}$) is less than 20 \%.
      }
  \end{center}
\end{figure}

In Fig.~\ref{fig_tc2}, we have plotted the experimental $T_c$ together with the $T_c$ calculated the most precisely in this work by including SOI and SF.
We note that since there is no computed data for Be and In with SOI, we are showing the data for them without SOI.
Excepting Cd, Zn, and V, we have accurately reproduced experimental $T_c$.
Inside the target elements of this benchmark, the three highest-$T_c$ elements observed experimentally are Nb (9.20 K), Tc (7.77 K), and Pb (7.19 K).
Also, the three highest-$T_c$ elements in our calculation are Nb (7.470 K), Tc (6.019 K), and Pb (6.010 K).
Therefore, at least in the elemental metals at ambient pressure, we can predict the highest-$T_c$ materials. 

\section{Discussion} \label{sec_discussion}

\begin{figure}[!tb]
  \begin{center}
    \includegraphics[width=8cm]{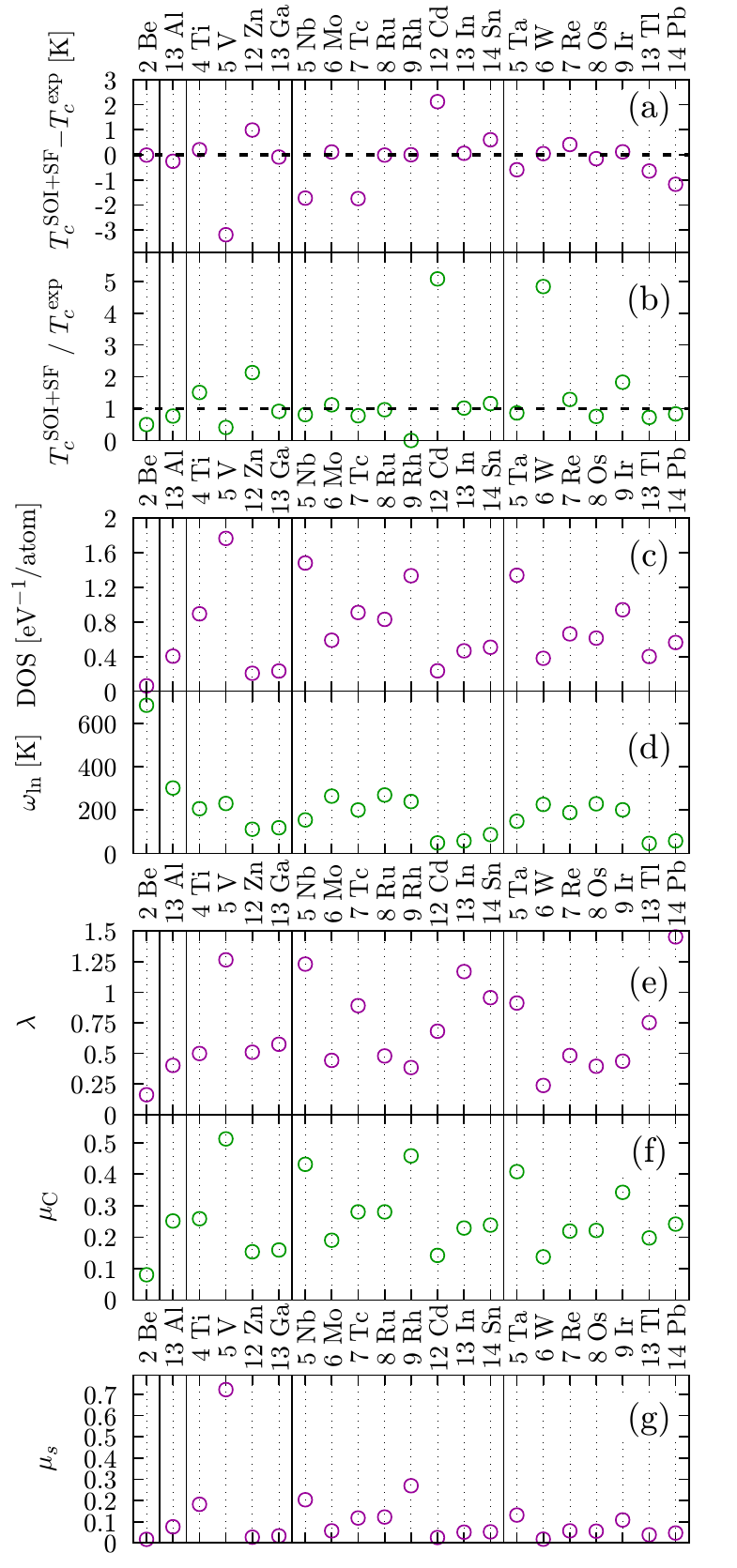}
    \caption{\label{fig_compare}
      We plot the following quantities for the elemental systems which have finite $T_c$:
      (a) The difference between the experimental $T_c$ ($T_c^{\rm exp}$) and
      theoretical $T_c$ computed with SOI and SF ($T_c^{\rm SOI+SF}$).
      (b) The ratio between $T_c^{\rm exp}$ and $T_c^{\rm SOI+SF}$.
      (c) The density of states at the Fermi level.
      (d) The averaged phonon frequencies $\omega_{\rm ln}$.
      (e) The Fr\"ohlich's mass-enhancement parameter $\lambda$.
      (f) The averaged Coulomb interaction $\mu_{\rm C}$ in Eq.~(\ref{eq_mu_c}).
      (g) The averaged SF term $\mu_s$ in Eq.~(\ref{eq_mu_s}).
      The horizontal axis is the atomic symbol together with the group of the periodic table.
      We note that since there is no computed data for Be and In with SOI,
      we show the data for them without SOI.
      }
  \end{center}
\end{figure}

In this section, we discuss the results shown in the previous section.
To check the accuracy of the calculation of the electron-phonon interaction,
we will plot in Fig.~\ref{fig_sommerfeld} the calculated- and experimental- \cite{GSCHNEIDNER1964275} Sommerfeld coefficient $\gamma_{\rm S}$
which is the prefactor of the specific heat at a low temperature
\begin{align}
  C_v = \gamma_{\rm S} T + O(T^3).
\end{align}
This coefficient is estimated using the density of states and the Fr\"ohlich's mass-enhancement parameter as follows
\begin{align}
  \gamma_{\rm S} = \frac{\pi^2 D(0)}{3} (1+\lambda).
\end{align}
The calculated $\gamma_{\rm S}$ agrees very well with the experimental value.
For Re, Pt, and Pb, this agreement becomes improved by including SOI;
Since these heavy elements have large SOI, this interaction is crucial to reproduce the experimental Sommerfeld coefficient.

We have plotted the computed- and experimental-$T_c$ data contained in Fig.~\ref{fig_periodic} into Fig.~\ref{fig_t_c}
to visualize the effect by SOI and SF;
we can detect the following trends by inspecting this graph:
SF always reduces $T_c$s for the elemental systems.
This reduction becomes significant for the transition metals and
is crucial to reproduce the experimental $T_c$ quantitatively.
The mechanism of this reduction is explained as follows \cite{PhysRevLett.17.433,tsutsumi}:
For isotropic superconductors such as elemental materials,
ferromagnetic spin-fluctuation becomes dominant and aligns the spin of electrons parallel.
This effect breaks the singlet Cooper pair in the isotropic superconductors.
On the other hand, in cuprates and iron-based superconductors,
highly anisotropic antiferromagnetic spin-fluctuation becomes dominant and
enhances the superconducting gaps with sign-changes \cite{PhysRevB.94.014503}.
In the transition metals, the effect of SF weakens with the increasing of the period number
in the periodic table.
For example, $\mu_s$ varies 0.722 (V) $\rightarrow$ 0.203 (Nb) $\rightarrow$ 0.131 (Ta),
0.057 (Mo) $\rightarrow$ 0.018 (W), 0.117 (Tc) $\rightarrow$ 0.057 (Re),
0.122 (Ru) $\rightarrow$ 0.055 (Os), and 0.269 (Rh) $\rightarrow$ 0.108 (Ir).
Also, the $\mu_s$ of Pd becomes negative; this indicates the formulation of SF
breaks down due to the magnetic order.
Although the magnetic order is a numerical artifact, this result shows that Pd has larger SF than that of Pt.
This trend of SF can be explained as follows \cite{tsutsumi}:
The electronic orbitals become delocalized with the increasing of the principal quantum number
(3d $\rightarrow$ 4d $\rightarrow$ 5d);
this delocalization decreases the magnetic exchange-correlation kernel in Eq.~(\ref{eq_xckernel});
also, the delocalized orbital has small DOS.
Therefore, the elements with the larger period number exhibit smaller SF contribution.
We cannot see this trend in the alkaline metals. For these elements,
$\mu_s$ does not decrease with the increasing of the period number, i.e.,
this parameter varies 0.213 (Na) $\rightarrow$ 0.270 (K) $\rightarrow$ 0.280 (Rb)
$\rightarrow$ 0.427 (Cs).
This behavior comes from the increasing of DOS because of the larger lattice constant
(larger atomic radius) for the alkaline metals having the larger period numbers.
The effect of SOI is small in most cases,
excepting Tc, Sn, Re, Tl, Pb.
In these elements, the Fr\"ohlich's parameter $\lambda$ changes drastically by turning on the SOI.
For Pb, this enhancement of $\lambda$ (1.036 $\rightarrow$ 1.453)
can be traced back to the three contributions i.e.,
the phonon softening ($\omega_{\rm ln}$ decreases from 67 K to 58 K),
the increased DOS at Fermi level (0.527 eV$^{-1}$ $\rightarrow$ 0.564 eV$^{-1}$), and
the enhanced deformation potential
${\boldsymbol \delta}^{{\bf q} \tau} V_{\rm KS}({\bf r})$ due to the
SOI term \cite{PhysRevB.81.174527}.
These effects of SOI in Tc, Re, and Tl are opposite to ones in Pb and Sn;
SOI reduces the electron-phonon coupling as well as $T_c$ in these three materials.
We can reproduce the absence of the superconductivity in
alkaline, alkaline earth, and noble metals, excepting Pt and Au with SOI and SF;
we have observed small finite $T_c$ for these two elements; we can reproduce the non-superconductivity also in Sc by including SF
while we observe $T_c=2.711$ K by ignoring SF.
Since Sc has highly localized 3d electrons, the SF largely reduces $T_c$.
For the group 12 elements (Zn and Cd), $T_c$s are overestimated
even if we include SF.
For these materials, the SF effect is small because the d orbitals are fully occupied.

Finally, we have tried to find the factor which dominates the accuracy of $T_c$.
In Fig.~\ref{fig_compare},
we have plottd the following quantities for the elemental systems with finite $T_c$:
(a) The difference between the experimental $T_c$ ($T_c^{\rm exp}$) and
theoretical $T_c$ computed with SOI and SF ($T_c^{\rm SOI+SF}$).
(b) The ratio between $T_c^{\rm exp}$ and $T_c^{\rm SOI+SF}$.
(c) The density of states at the Fermi level.
(d) The averaged phonon frequencies $\omega_{\rm ln}$ in Eq.~(\ref{eq_omegaln}).
(e) The Fr\"ohlich's mass-enhancement parameter $\lambda$ in Eq.~(\ref{eq_frohlich}).
(f) The averaged Coulomb interaction $\mu_{\rm C}$ in Eq.~(\ref{eq_mu_c}).
(g) The averaged SF term $\mu_s$ in Eq.~(\ref{eq_mu_s}).
Note that since there is no computed data for Be and In with SOI,
we are presenting data for these two elements without SOI.
$T_c$s of Zn and Cd (V, Nb, Tc, Pb) are overestimated (underestimated)
in the differential plot (a)
while those of Zn Cd, W, Ir (Be, V, Rh) are overestimated (underestimated)
in the ratio plot (b).
Therefore, for V, Zn, and Cd, the focus should be on examining the accuracy of $T_c$; therefore, we attempted to find features of these three materials from elemental systems.
From Fig.~\ref{fig_compare}~(g), we can see that V has extremely large $\mu_s$.
When the system has large SF, the SF-mediated interaction in Eq.~(\ref{eq_hatlambdasf_fourier}) changes rapidly because the inverse of the matrix in Eq.~(\ref{eq_stoner_criterion}) approaches to a singular matrix.
Therefore, the SF of such systems needs to computed more precisely, for example, by including the gradient collection into the magnetic exchange-correlation kernel in Eq.~(\ref{eq_xckernel}).
However, it is difficult to identify the significant difference between Zn, Cd, and other materials.
For example, the parameters of Ga are extremely close to those of Zn. However, the calculated $T_c$ of Ga is in good agreement with the experimental one.

By leaving from the superconductivity, we can see the following features from Fig.~\ref{fig_compare}:
DOS and $\mu_C$ are showing very similar behavior.
Since $\mu_C$ in Eq.~(\ref{eq_mu_c}) can be approximated to the DOS times the averaged Coulomb interaction, the synchronicity indicates that the elemental materials in Fig.~\ref{fig_compare} have nearly the same screened Coulomb repulsion.
$\omega_{\rm ln}$ has peaks around group 6-9 on both periods 5 and 6.
Additionally, the frequencies at both the peaks are extremely close, although the atomic masses of these periods are different;
these behaviors can be explained by the following Friedel's theory \cite{friedel-model}.
The materials in groups 6-9 have high cohesive energy because of the half-filled d-orbitals; the high cohesive energy leads to the hardness and the high phonon frequencies of the materials.
Since 5d orbitals are more delocalized than 4d orbitals, the binding energy increases for 5d materials.
Therefore, the phonon frequency is unchanged because of the cancellation between the atomic mass and the stronger bonding.

\section{Summary} \label{sec_summary}

We performed the benchmark calculations of SCDFT using our open-source software package {\sc Superconducting Toolkit}, which uses our newly developed method for treating SOI together with SF.
We presented benchmark results of superconducting properties
calculated by SCDFT for 35 elemental materials together with computational details,
and discussed the accuracy of the predicted $T_c$
and the effects of SF and SOI up on $T_c$.
We found that the calculations, including SOI and SF, can quantitatively reproduce the experimental $T_c$s.
The SF is essential especially for the transition metals; still
the effect of SOI is small for elemental systems excepting Tc, Sn, Re, Tl, and Pb.
We also reproduced the absence of the superconductivity in the alkaline, alkaline earth, and noble metals.
Our result could be used to check the validity of future work
such as the high-throughput calculation for exploring new superconductors.
Moreover, the knowledge of this benchmark can be used to improve the methodology
of SCDFT.
For example, we can focus on Zn and Cd as a target for the next theoretical improvement.
It is straightforward to extend the current benchmark calculation into the binary, ternary, and quaternary systems.
From such a benchmark, we check systematically the accuracy of SCDFT for the compound superconductors such as magnesium diboride \cite{ISI:000167194300040}, cuprates \cite{ISI:000282019500009}, iron-based \cite{RevModPhys.83.1589}, and heavy-fermion superconductors \cite{RevModPhys.81.1551}, etc.
This type of benchmark calculation should be performed whenever there is room for theoretical improvements so that the applicability of SCDFT can be extended as a universal tool.

\begin{acknowledgments}
  We thank Ryosuke Akashi for the fruitful discussion about the spin-fluctuation. 
  This work was supported by 
  Priority Issue (creation of new functional devices and high-performance materials
  to support next-generation industries)
  to be tackled using Post `K' Computer
  from the MEXT of Japan.
  Y. H. is supported by Japan Society for the Promotion of Science through Program for Leading Graduate Schools (MERIT).
  The numerical calculations in this paper were done on the supercomputers in ISSP and Information Technology Center at the University of Tokyo.
\end{acknowledgments}


\bibliography{ref}

\end{document}